\documentclass[aip,reprint,amsmath,amssymb]{revtex4-1}

\usepackage{graphicx}
\usepackage{dcolumn}
\usepackage{float}
\usepackage{bm}
\usepackage{amssymb}
\usepackage{lipsum}

\begin{document}

\title{Equilibrium organization, conformation, and dynamics of two polymers 
under box-like confinement}

\author{James M. Polson and Desiree A. Rehel}
\affiliation{ Department of Physics, University of Prince Edward Island,
550 University Avenue, Charlottetown, Prince Edward Island, C1A 4P3, Canada }

\date{\today}

\begin{abstract}
{Motivated by recent nanofluidics experiments}, we use
Brownian dynamics and Monte Carlo simulations to study the conformation, organization and dynamics 
of two polymer chains confined to a single box-like cavity. {The polymers are modeled 
as flexible hard-sphere chains, and} the box has a square cross-section of side length $L$
and a height that is small enough to compress the polymers in that dimension. {For 
sufficiently large $L$}, the system behaviour approaches that of an isolated polymer in a slit. 
However, the combined effects of crowding and confinement {on the polymer organization, 
conformation and equilibrium dynamics} become significant when $L/R_{{\rm g},xy}^*\lesssim 5$, 
where $R_{{\rm g},xy}^*$ is the transverse radius of gyration for a slit geometry. In this regime, 
{the centre-of-mass probability distribution in the transverse plane exhibits a depletion 
zone near the centre of the cavity (except at very small $L$) and a 4-fold symmetry with
quasi-discrete positions.} Reduction in polymer size with decreasing $L$ arises principally 
from confinement rather than {inter-polymer} crowding. By contrast, polymer diffusion 
and internal motion are strongly affected by {inter-polymer} crowding. {
The two polymers tend to occupy opposite positions relative to the box centre, about which they 
diffuse relatively freely. Qualitatively, this static and dynamical behaviour differs significantly 
from that previously observed for confinement of two polymers to a narrow channel.} The simulation 
results for a suitably chosen box width are qualitatively consistent with results from a recent 
experimental study of two $\lambda$-DNA chains confined to a nanofluidic cavity.
\end{abstract}

\maketitle

\section{Introduction}
\label{sec:intro}

Nanoscale confinement of a polymer strongly affects its conformational and dynamical 
properties.  Recent advances in nanofabrication techniques have facilitated the creation of 
lab-on-a-chip nanofluidic devices that are ideal for studying and characterizing 
such physical effects.  In recent years, nanofluidics experiments employing optical 
imaging techniques to study biopolymers such as DNA have been instrumental in testing and refining 
decades-old theories of confined polymers. A notable example is confinement of DNA in 
nanochannels.\cite{dai2016polymer,reisner2012dna} A thorough understanding of the 
fundamental physics of such systems is vital for various applications that require 
stretching of DNA in channels, including DNA sorting,\cite{dorfman2012beyond} 
DNA denaturation mapping,\cite{reisner2010single,marie2013integrated} and genome mapping.%
\cite{lam2012genome,hastie2013rapid,dorfman2013fluid,muller2017optical} 
A number of other studies have examined the confinement effects on DNA in embedded nanotopography
devices composed of a nanoslit with nanogrooves or nanopits etched into one surface
deeper than the surrounding slit. While the pits and grooves generally promote entropic trapping 
of the DNA, some portion of the contour of the molecule can occupy the narrow region
of the slit outside these structures. This enables a novel means for conformational manipulation 
of single polymers. For example, recent experiments by Reisner and coworkers have observed 
and characterized `digitized' or `tetris'-like conformations for polymers that share their 
contour between multiple adjacent nanopits.\cite{klotz2015correlated,klotz2015measuring} 
Embedded nanotopography devices can also serve as useful models to characterize single-molecule 
transport across free energy landscapes.\cite{mikkelsen2011pressure,kim2017giant,%
smith2017photothermal,klotz2016waves,del2009pressure,klotz2012diffusion,ruggeri2017lattice}

One notable study using nanofluidics techniques to study the effects of nanoconfinement on polymers 
was reported recently by Capaldi {\it et al.}\cite{capaldi2018probing} Their experiment 
employed pneumatic pressure to deflect a thin nitride lid into a nanoslit containing a solution
of fluorescently stained $\lambda$-DNA chains, forcing the molecules into an array of nanocavities 
embedded in one surface of the slit. Each cavity had a square cross section of side length 2~$\mu$m,
was 200~nm deep, and was able to trap up to two $\lambda$-DNA chains per cavity.  
Differential staining of the chains enabled monitoring of 
individual chain conformation, the degree of partitioning or mixing of the chains, and coupled
diffusion of the centre-of-mass chain positions. Comparing the results to those for cavities
with a single trapped DNA chain, the drastic impact of the presence of a second chain
on the conformation and dynamics was quantified. Similar, though less pronounced, effects
were observed for a cavity-confined system containing a single $\lambda$-DNA chain together 
with a small plasmid. 

Numerous theoretical and simulation studies have examined the mixing/partitioning behaviour 
of two polymers confined to nanoscale cavities and channels in recent years.\cite{jun2006entropy,%
teraoka2004computer,jun2007confined,arnold2007time,jacobsen2010demixing,jung2010overlapping,%
jung2012ring,jung2012intrachain,liu2012segregation,dorier2013modelling,racko2013segregation,%
shin2014mixing,minina2014induction,minina2015entropic,chen2015polymer,polson2014polymer,%
du2018polymer,polson2018segregation,nowicki2019segregation,nowicki2019electrostatic,polson2021free} 
Under sufficiently strong confinement, polymers tend to segregate due to entropic repulsion between
the chains.  It has been suggested that this effect may contribute to the driving force for chromosome
segregation in self-replicating bacteria,\cite{jun2006entropy,jun2010entropy,diventura2013chromosome,%
youngren2014multifork,mannik2016role} and recent experimental studies have reported results consistent
with this hypothesis.\cite{diventura2013chromosome,mannik2016role,cass2016escherichia,wu2020geometric,%
elnajjar2020chromosome,japaridze2020direct} 
Unfortunately, {\it in vivo} experiments on replicating bacteria do not provide a straightforward
means to quantify the degree of entropic repulsion. By contrast, {\it in vitro} nanofluidics experiments 
such as that by Capaldi {\it et al.}, which consider much simpler systems, are much better
suited for direct comparison with the predictions of theory and simulation.

In this study, we use Brownian dynamics (BD) and Monte Carlo (MC) simulations to study the organization,
conformational behaviour, and equilibrium dynamics of a system of two polymers under confinement
in a box-like cavity. Ideally, the molecular model should incorporate correct length scale ratios 
for the width, contour length, and persistence length of $\lambda$-DNA. However, this choice leads 
to simulations that are far too time consuming to be of practical benefit, especially in the case 
of dynamics.  Consequently, we employ instead a simple coarse-grained molecular model, in which the 
polymers are described as relatively short flexible chains of spherical Lennard-Jones beads. One goal 
of this study is to determine whether the general trends observed in the study of Capaldi {\it et al.} 
can be accounted for using such a simplistic model. The simulations also provide a means to
test the validity of the interpretation proposed by Capaldi {\it et al.} for the observed dynamical
behaviour. In addition, we examine effects of varying system parameters not considered in the 
experiments. Most notably, we study the effects of varying the confining box dimensions on the 
polymer dynamics and organization. For sufficiently small cavities, we find that the polymers tend 
to segregate to opposite sides of the box and that the rates of polymer diffusion and internal 
motion are both strongly affected by interpolymer crowding. These observations are qualitatively
consistent with those of the experimental study, demonstrating the utility of the very simplistic
model employed in the simulations.  The observed behaviour in this model system may also be of value in
interpreting results of future experiments.

The remainder of this article is organized as follows. Section~\ref{sec:model} presents a
brief description of the model used in the simulations, following which
Section~\ref{sec:methods} gives an outline of the methodology employed together with
the relevant details of the simulations. Section~\ref{sec:results} presents the simulation
results for the various systems we have examined, and Section~\ref{sec:experiment} describes
the relevance of the simulation results to experiment.  Finally, Section~\ref{sec:conclusions}
summarizes the main conclusions of this work.

\section{Model}
\label{sec:model}

We examine systems of either one or two polymer chains confined to a box-like cavity.
Each polymer is a flexible linear chain of $N$ spherical monomers. Polymer lengths are in
the range $N$=40--80 for BD simulations and 40--300 for MC simulations. {
For the two-polymer systems, the lengths of the two polymers are equal.}
Non-bonded interactions are given by the repulsive Lennard-Jones potential,
\begin{eqnarray}
u_{\rm nb}(r) =
\begin{cases}
 u_{\rm LJ}(r) - u_{\rm LJ}(r_{\rm c}),  & r \leq r_{\rm c} \\
0,   & r \geq r_{\rm c}
\end{cases}
\label{eq:LJ}
\end{eqnarray}
where $r$ is the distance between the monomer centres, $r_{\rm c} \equiv 2^{1/6}\sigma$,
and where $u_{\rm LJ}(r)$ is the standard Lennard-Jones 6-12 potential,
\begin{equation}
u_{\rm LJ}(r) = 4\epsilon\left[\left(\frac{\sigma}{r}\right)^{12}-
\left(\frac{\sigma}{r}\right)^6\right].
\end{equation}
Bonded monomers interact with a combination of the potential in Eq.~(\ref{eq:LJ})
and the finite extensible nonlinear elastic (FENE) potential,
\begin{equation}
u_{\rm FENE}(r)=-{\textstyle\frac{1}{2}} kr_0^2 \ln(1-(r/r_0)^2), 
\label{eq:FENE}
\end{equation}
where we choose $k\sigma^2/\epsilon=30$ and where $r_0=1.5\sigma$.

The polymers are enclosed in a rectangular box with a square cross section in the $x-y$ 
plane of side length $L$ and a height $h$ in the $z$ direction. The box dimensions are 
defined such that $L$ is the range of positions along the $x$ and $y$ axes accessible to 
the centres of the monomers, and $h$ is the corresponding range along $z$. To impose
this condition, each monomer interacts with each wall through Eq.~(\ref{eq:LJ}), where 
$r+\sigma$ is the distance of the monomer to the nearest point on the wall.  Most 
calculations used $h$=4, a value that is low enough to compress the polymer along the
$z$ direction, as was the case in the experiments of Ref.~\onlinecite{capaldi2018probing}.  
We use a wide range of values for the box width. 

\section{Methods}
\label{sec:methods}

We use two different simulation methods to study the confined-polymer system. BD simulations 
are used to monitor the dynamics of centre of mass motion as well as the internal dynamics 
of each chain. {Since the focus is on characterizing dynamics at longer time 
scales, results obtained using the BD method are not expected to differ significantly from 
those obtained using the more computationally costly Langevin dynamics method.} 
MC simulations employing the standard Metropolis method are used to measure 
probability distributions associated with polymer position, as well as to characterize the 
conformational statistics.  {Although in principle BD simulations could also be 
used for these measurements, they were far too computationally costly for larger $N$ to
obtain statistically sound results in a reasonable time. By contrast, this presented
no problem for the much more efficient MC simulations. For convenience, we chose to
use MC simulations for the static quantities for all $N$.}  Both methods employ the molecular 
model described in Section~\ref{sec:model}. A brief description of each method follows below.

\subsection{Monte Carlo simulations}
\label{subsec:mc}

We use a standard MC simulation method in which polymer configurations are generated using 
trial moves that are accepted or rejected based on the Metropolis MC criterion. The trial
moves consist of a combination of single-monomer crankshaft rotations, reptation 
moves and whole-polymer displacements. The type of each trial move is randomly selected.
A single MC cycle consists of $2N$ trial moves,  each consisting
of of $N-1$ crankshaft moves, $N-1$ reptation moves, and two whole-polymer translations. 
For each crankshaft move, a randomly selected monomer was rotated about an 
axis connecting adjacent monomers through a random angle drawn from uniform distribution in 
the range $[-\Delta \phi_{\rm max}, +\Delta\phi_{\rm max}]$. In the case of end monomers, 
rotation was about the second bond from the end.  Whole-polymer translation 
was achieved by moving all monomers of a randomly selected polymer through a displacement drawn 
from a uniform distribution in the range $[-\Delta_{\rm max},+\Delta_{\rm max}]$ for each 
coordinate. The parameters $\Delta \phi_{\rm max}$ and $\Delta_{\rm max}$ were chosen to achieve 
an acceptance ratio of approximately 50\%. For each reptation move, the polymer and the
reptation direction were both randomly selected.

{For each system size, as defined by $N$ and $L$, we carried out numerous simulations on
an array of processors, each using a different sequence of random numbers, to acquire a collection 
of statistically uncorrelated results. This collection of results was then averaged. Dividing the 
calculation into such independent runs essentially parallelizes the simulation and dramatically
increases the computational efficiency.} Each simulation consisted of an equilibration period of 
typically $10^6$ MC cycles followed by a production run of $10^8$ MC cycles. The number of these 
simulations ranged from 50 for $N$=40 to 1000 for $N$=300, corresponding to total simulation times 
of 540 CPU-hours for $N$=40 to 15000 CPU-hours for $N$=300, respectively.

\subsection{Brownian dynamics simulations}
\label{subsec:bd}

The BD simulations used to study the polymer dynamics employ standard methods.
The coordinates of the {\it i}th particle are advanced through a time $\Delta t$ according to 
the algorithm:
\begin{eqnarray}
x_i(t+\Delta t) & = & x_i(t) + \frac{f_{i,x}}{\gamma_0} + \sqrt{2 k_{\rm B}T \Delta t/\gamma_0} \Delta w,
\label{eq:BDeq}
\end{eqnarray}
and likewise for $y_i$ and $z_i$.  Here, $f_{i,x}$ is the $x$-component of the conservative 
force on particle $i$, and $\gamma_0$ is the friction coefficient of each monomer. The conservative 
force is calculated as $f_{i,\alpha}=-\nabla_{i,\alpha} U_{\rm tot}$, where $\nabla_{i,\alpha}$ 
is the $\alpha$-component of the gradient with respect to the coordinates of the $i$th particle 
of the total potential energy of the system, $U_{\rm tot}$. In addition, $\Delta w$ is a random 
quantity drawn from a Gaussian of unit variance.  

All simulations were carried out at a temperature $T = \epsilon/k_{\rm B}$, where $k_{\rm B}$
is Boltzmann's constant.  The time step used in Eq.~(\ref{eq:BDeq}) was 
$\Delta t = 0.0001\tau_{\rm BD}$, where $\tau_{\rm BD}\equiv\gamma_0\sigma^2/\epsilon$. The 
run time of each simulation was typically $10^5\tau_{\rm BD}$, following an equilibration 
period of typically $10^4\tau_{\rm BD}$. For each system size, as defined by $N$ and $L$, the 
results of numerous simulations were averaged and used to estimate uncertainties. This ranged from 
750 simulations for $N$=40 to 1000 simulations for $N$=80, corresponding to total simulation times 
of roughly 10000 CPU-hours and 30000 CPU-hours, respectively.

\subsection{Measured quantities}
\label{subsec:measurements}

To track the polymer centre-of-mass motion, we use the mean-square displacement, 
\begin{eqnarray}
{\rm MSD}(t) = \left\langle \left( x_i(t) - x_i(0) \right)^2 \right\rangle.
\label{eq:MSDdef}
\end{eqnarray}
where $x_i$ is the centre-of-mass $x$ coordinate for the $i$th polymer. The angular brackets 
denote an average over sequences of configurations generated in separate simulations, as 
well as over the time origin for each simulation. In addition, an average was carried out 
over both polymer positions in the 2-polymer system, as well as over the $y$ coordinates 
of the centres of mass. The latter average is valid since the box width in the $x$ and $y$ 
dimensions is equal. 

A related measure of polymer motion is the position autocorrelation function,
\begin{eqnarray}
C_{\rm auto}(t) \equiv \left\langle x_i(t)x_i(0)\right\rangle.
\label{eq:Cauto}
\end{eqnarray}
Note that since the box centre lies at $x=0$, then $\langle x_i\rangle=0$, and
thus $\langle  x_i(t)x_i(0)\rangle = \langle x_i^2\rangle - \frac{1}{2}{\rm MSD}(t)$.
Correlations between the centre-of-mass kinetics of the two polymers is quantified
by the cross-correlation function,
\begin{eqnarray}
C_{\rm cross}(t) \equiv -\left\langle x_1(t)x_2(0)\right\rangle,
\label{eq:Ccross}
\end{eqnarray}
where the subscripts denote different polymers. Since the signs of $x_1$ and $x_2$ tend 
to be opposite (i.e. the polymers tend to be situated on opposite sides of the box), the 
negative sign leads to the property, $C_{\rm cross}(t) \geq 0$.

To examine internal motion of the polymers, we use Rouse coordinates, defined as
\begin{eqnarray}
{\bf R}_p \equiv \frac{1}{N}\sum_{n=1}^N {\bf r}_n \cos\left(\frac{p(n-{\textstyle\frac{1}{2}})\pi}{N}\right),
\end{eqnarray}
where ${\bf r}_n$ is the position of particle $n$, and $p=1,2,3...$. These are used to calculate the
correlation functions
\begin{eqnarray}
C_p(t) & = & \left\langle {\bf R}_p(t+t_0)\cdot {\bf R}_p(t_0)\right\rangle_{xy},
\end{eqnarray}
where the subscript indicates that only transverse components of the coordinates are used in
the calculation of the average. In most cases we find that the correlation function decays 
exponentially such that $C_p\propto e^{-t/\tau_p}$, where $\tau_p$ is the correlation 
time for the $p$th mode. Typically, we observe a small transient at short times, which is excluded
from the fit. {The consistently exponential form of $C_p(t)$ for both one- and 
two-polymer systems with excluded volume under even strong confinement conditions is a somewhat
surprising but useful property. For example,} we find that the $p=1$ mode provides a more 
convenient probe of large-scale conformational changes than, e.g., the end-to-end displacement 
since the correlation function of the latter tends not to be exponential under the conditions 
examined here.

As a measurement of the polymer shape isometry, we use the 2-D version of asphericity,
$A_2$, defined as
\begin{eqnarray}
A_2 = \frac{\left\langle R_1^2\right\rangle - \left\langle R_2^2\right\rangle } 
         {\left\langle R_1^2\right\rangle + \left\langle R_2^2\right\rangle },
\label{eq:asph}
\end{eqnarray}
where the angular brackets denote an average over sampled configurations.
The quantities $R_1^2$ and $R_2^2$ ($\leq R_1^2$) are eigenvalues of the 
2-D gyration matrix, whose elements are defined
\begin{eqnarray}
S_{\alpha\beta} & = & \frac{1}{N} \sum_{i=1}^{N} 
\left(r_{\alpha,i} - r_{\alpha,{\rm cm}}\right)
\left(r_{\beta,i}  - r_{\beta,{\rm cm}}\right), \\
&&\nonumber
\end{eqnarray}
where $r_{\alpha,i}$ is the $\alpha$-coordinate ($\alpha=x, y$) of particle $i$ and 
$r_{\alpha,{\rm cm}}$ is the instantaneous $\alpha$-coordinate of the centre-of-mass.  The 
quantities $R_1$ and $R_2$ ($\leq R_1$) can be viewed as the semi-major and 
semi-minor axes of an equivalent ellipse that very roughly approximates the shape
of the polymer in the $x-y$ plane. Note the two limiting cases for the asphericity: 
$A_2=0$ corresponds to a circular disk, and $A_2=1$ corresponds to an infinitesimally thin
needle. Note as well that $R_{{\rm g},xy}^2 = R_1^2+R_2^2$ is the instantaneous
square radius of gyration in the $x-y$ plane.
As a measure of inter-polymer overlap, we define the overlap parameter $\chi_{\rm ov}\equiv 
N_{\rm ov }/N$, where $N_{\rm ov}$ is the average number of monomers per polymer that lie 
inside the overlap area of the two equivalent ellipses defined above. 


In the results presented below, distances are measured in units of $\sigma$, energy
is measured in units of $\epsilon$ ($=k_{\rm B}T$), and time is measured in units of 
$\gamma_0\sigma^2/\epsilon$.

\section{Results}
\label{sec:results}

\subsection{Polymer organization and conformational statistics}
\label{subsec:statics}

\begin{figure*}[!ht]
\begin{center}
\vspace*{0.2in}
\includegraphics[angle=0,width=0.95\textwidth]{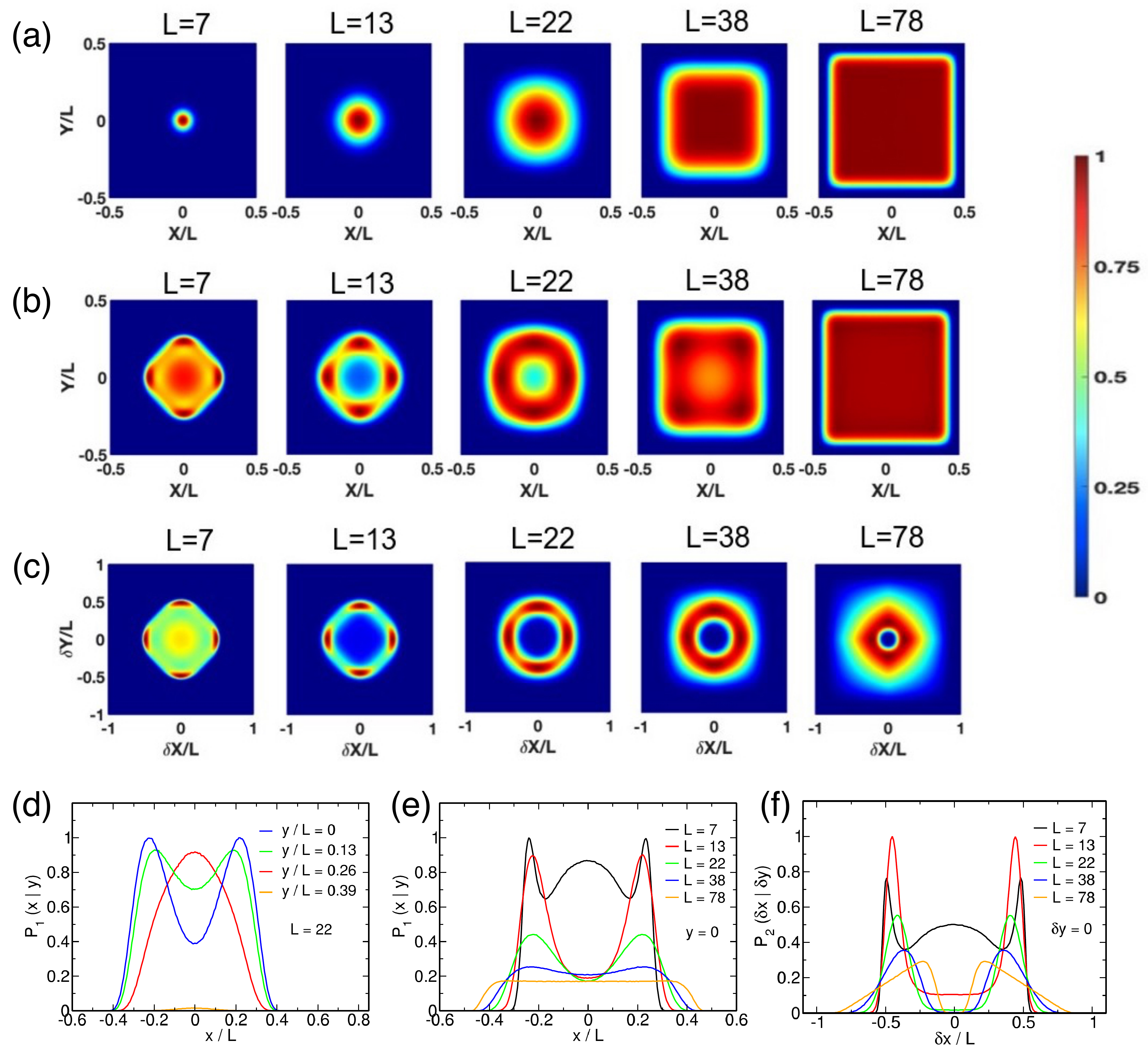}
\end{center}
\caption{Probability distributions for a system of one and two polymers of length $N$=60 confined to a 
box of height $h$=4. Results for various box widths are shown. Row (a) shows results for the distribution
${\cal P}_1(x,y)$ for a {\it single-polymer system}. Row (b) shows distribution ${\cal P}_1(x,y)$ for
a {\it two-polymer system}. Row (c) shows ${\cal P}_2(\delta x,\delta y)$, the probability
distributions for the {\it difference} in the centre-of-mass coordinates, $\delta x$ and $\delta y$
of the polymers in the two-polymer system. {In (a), (b) and (c) the axes are scaled by 
the box width, $L$, and the colour intensity maximum corresponds to the maximum value of the probability 
for each graph.  (d) Probability distribution cross-sections ${\cal P}_1(x|y)$ for the two-polymer system 
with $L$=22 for various values of $y$. (e) As in (d), except for $y$=0 (which bisects the box) for various 
$L$.  (f) Probability distribution cross-sections ${\cal P}_2(\delta x|\delta y)$ for $\delta y$=0 and 
for various $L$.}}
\label{fig:prob2d_N60}
\end{figure*}

We consider first the effects of confinement and crowding on the organization of the polymers 
in the cavity.  Figure~\ref{fig:prob2d_N60} shows 2-D probability distributions that characterize 
the position of the polymers inside the the box-like cavity in the transverse plane. Results
are shown for wide range of box widths.  {The lowest value of $L$ was chosen to 
maintain the condition $L>h$, while the highest value of $L$ (as explained below) corresponds
to a system for which confinement and inter-polymer crowding effects are negligible.}
{Row (a)} shows the probability of the centre
of mass of a polymer at any position in the $x-y$ plane for the case where just a single
polymer is confined to the cavity. {Row (b)} shows the same probability distribution
for the case where two identical polymers are confined to the box.  We label these distributions
${\cal P}_1(x,y)$, where the subscript denotes the single-polymer aspect of the distribution
(and {\it not} the number of confined polymers). {Row (c)} shows probability distributions 
for the difference in the centre-of-mass coordinates, $\delta x$ and $\delta y$, of the polymers 
in the two-polymer system. We label these distributions ${\cal P}_2(\delta x,\delta y)$, where 
the subscript denotes the fact that this is a 2-polymer property. (Note that this quantity
is a physically meaningful descriptor only for the 2-polymer system.)  Together, the two complementary 
distributions provide a clear picture of the effect of varying box width $L$ on the polymer 
position. {As an aid to interpret these 2D probability maps, Figs.~\ref{fig:prob2d_N60}(d), 
(e) and (f) show probability cross sections of ${\cal P}_1(x,y)$ and ${\cal P}_2(\delta x,\delta y)$ 
for some of the two-polymer systems.}

A useful relative measure of the box width is the ratio $L/R_{{\rm g},xy}^*$, where 
$R_{{\rm g},xy}^*$ is the transverse root-mean-square radius of gyration for a polymer in a 
slit, i.e., $L$=$\infty$.  A box with $L/R_{{\rm g},xy}^*\gg 1$ is large in the sense that 
the polymers are unlikely to interact with either the walls or with each other. As the box 
width approaches the regime where $L/R_{{\rm g},xy}^*$ is of the order of unity, the combined 
effects of the polymer-polymer and polymer-wall interactions are expected to strongly affect 
the organization of the polymers inside the box as well as their conformational behaviour.  
Note that the system in Fig.~\ref{fig:prob2d_N60} is characterized by $N$=60 and $h$=4, for 
which $R_{{\rm g},xy}^*=5.710\pm 0.001$.

Let us consider first the behaviour of the 1-polymer system. At the largest box size of $L=78$,
the probability distribution ${\cal P}_1(x,y)$ is fairly flat over the $x-y$ plane, with
the exception of an entropy-induced depletion layer along the lateral walls. As $L$ is reduced,
relative width of the depletion zone grows, and the polymer centre-of-mass distribution 
increasingly narrows to the box centre. Concomitantly, the distribution transforms from a
square shape (corresponding to the shape of the box) to a circular shape.

A comparison of the ${\cal P}_1(x,y)$ distributions for the 1-polymer system with those for
the 2-polymer systems shows the increasingly pronounced effect of crowding as $L$ decreases.
For the largest box size of $L=78$ ($L/R_{{\rm g},xy}^*=13.7$), ${\cal P}_1(x,y)$ is essentially
identical to that for the 1-polymer system: flat over the $x-y$ plane, with an entropy-induced 
depletion layer along the lateral walls. This is precisely the behaviour expected in the dilute 
limit where interpolymer interactions are infrequent. The corresponding distribution 
${\cal P}_2(\delta x,\delta y)$ exhibits a high-probability ring with a probability hole in the 
middle, as well as very low probability for large inter-polymer displacements. The depletion hole 
corresponds simply to the tendency of the polymers to avoid transverse overlap, as such configurations 
would reduce the conformational entropy. Likewise, very large separations require the polymers to 
press against the walls of the box, which also has an entropy-reducing effect. Within the 
high-probability ring, there are slight enhancements at four symmetrically distributed locations, 
two at $\delta x=0$ and two at $\delta y=0$. 

As $L$ becomes smaller, ${\cal P}_1(x,y)$ changes significantly. At $L$=38 ($L/R_{{\rm g},xy}^*$=6.7),
small peaks appear near the corners of the distribution. Thus, the polymers increasingly tend to 
be situated near the corners of the box. No such feature is present for the 1-polymer system,
indicating that it is a consequence of interpolymer crowding.  At a smaller box size 
$L$=22 ($L/R_{{\rm g},xy}^*$=3.9), the distribution has a ring-like structure, with a significant
depletion hole in the centre of the box. Slight enhancements are evident at four symmetrically
related positions. The narrowness of the ring structure for the corresponding
${\cal P}_2(\delta x,\delta y)$ indicates that when one polymer centre lies at $(x,y)$ the other
will tend to lie at an inverted position of $(-x,-y)$. Thus, each polymer centre is expected to 
diffuse around the centre of the box with the other polymer moving in a correlated manner in
this inverted position.  Note the pronounced qualitative difference in ${\cal P}_1(x,y)$ between
the one- and two-polymer systems.

As the box size decreases to $L$=13 ($L/R_{{\rm g},xy}^*$=2.3) the ring structure for the 
two distributions gives way to strong probability enhancements at four symmetrically related 
positions. For ${\cal P}_1$, two are located along $x$=0 and the other two are at $y$=0. 
A similar structure is evident for ${\cal P}_2$.
This behaviour indicates that the polymers tend to occupy quasi-discrete locations 
in opposite halves of the box divided by boundaries at $x$=0 or $y$=0. Such behaviour is a consequence 
of the significant interaction between the polymers (i.e., crowding) and between each polymer and 
the confining walls (i.e., confinement) for this small box size. As will be shown below, these 
interactions also strongly deform the polymer conformation leading to significant changes 
in its average size and shape.  At the smallest examined box size of $L=7$ ($L/R_{{\rm g},xy}^*$=1.23), 
a new trend emerges: an enhancement of the probability at positions near the centre of the box 
and for very small inter-polymer displacements (i.e. near $\delta x = \delta y = 0$).  Note that 
these broad probability peaks for ${\cal P}_1(x,y)$ and ${\cal P}_2(\delta x,\delta y)$ each coexist 
with the remaining four peaks associated with the the quasi-discrete states described above.
Again, we note the qualitatively different behaviour in ${\cal P}_1(x,y)$ for the 1- and 2-polymer
systems for these small box sizes.

Although the distributions in Fig.~\ref{fig:prob2d_N60} were calculated for $N$=60 and $h$=4,
the qualitative trends are unaffected by other choices of $N$ and $h$ (as long as $h$ is
sufficiently small to compress the polymer along the $z$ direction). Distributions for a polymer
length of $N$=300 and box heights of $h$=4, 6, and 8 presented in Figs.~1 and 2 of the ESI indeed 
show the same trends as those in Fig.~~\ref{fig:prob2d_N60} above.\dag\footnotetext{\dag~Electronic 
Supplementary Information (ESI) available: [details of any supplementary information available 
should be included here]. See DOI: 10.1039/b000000x/}

\begin{figure}[!ht]
\begin{center}
\vspace*{0.2in}
\includegraphics[width=0.4\textwidth]{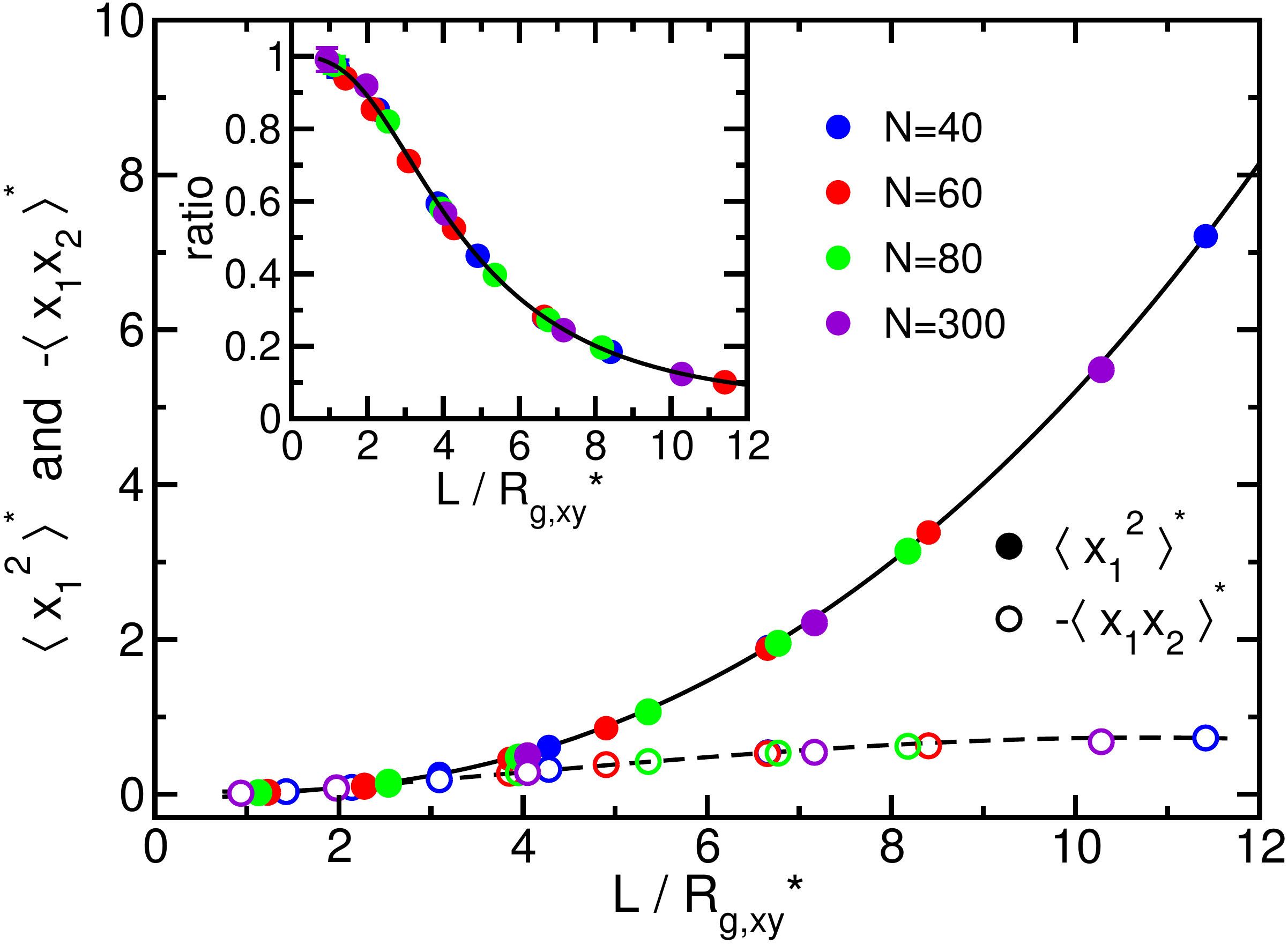}
\end{center}
\caption{Variation of $\langle x_1^2\rangle^*\equiv \langle x_1^2\rangle/(R_{{\rm g},xy}^*)^2$ 
(closed symbols) and $-\langle x_1 x_2\rangle^*\equiv -\langle x_1 x_2\rangle/(R_{{\rm g},xy}^*)^2$ 
(open symbols) with scaled box width, $L/R_{{\rm g},xy}^*$. Results are shown for $N$=40, 60, 80, 
and 300. The solid and dashed curves are guides for the eye. The inset shows the ratio of the two 
quantities for each chain length.}
\label{fig:corr_LJ}
\end{figure}

The tendency of the polymer positions to become inversely correlated with respect to the box
centre for highly confined systems is also illustrated in Fig.~\ref{fig:corr_LJ}. 
{The mean-square centre-of-mass position of each polymer, $\langle x_1^2\rangle$ 
($=\langle x_2^2\rangle$),
and the position cross-correlation, $-\langle x_1 x_2\rangle$, both decrease monotonically
with decreasing $L$.} However, the ratio, $-\langle x_1x_2\rangle/\langle x_1^2\rangle$, shown 
in the inset increases as the polymers become more confined. This ratio is a measure of the
degree of inverse correlation of the polymer positions. Note that the ratio is independent
of $N$ when plotted against the scaled box length, $L/R_{{\rm g},xy}^*$. We expect qualitatively 
similar behaviour for more a realistic polymer model (e.g., one providing a better description of 
$\lambda$ DNA), except with a much lower degree of correlation. The convergence of the ratio to 
unity here likely arises from packing effects due to the high volume fraction at low $L$.

\begin{figure}[!ht]
\begin{center}
\vspace*{0.2in}
\includegraphics[width=0.45\textwidth]{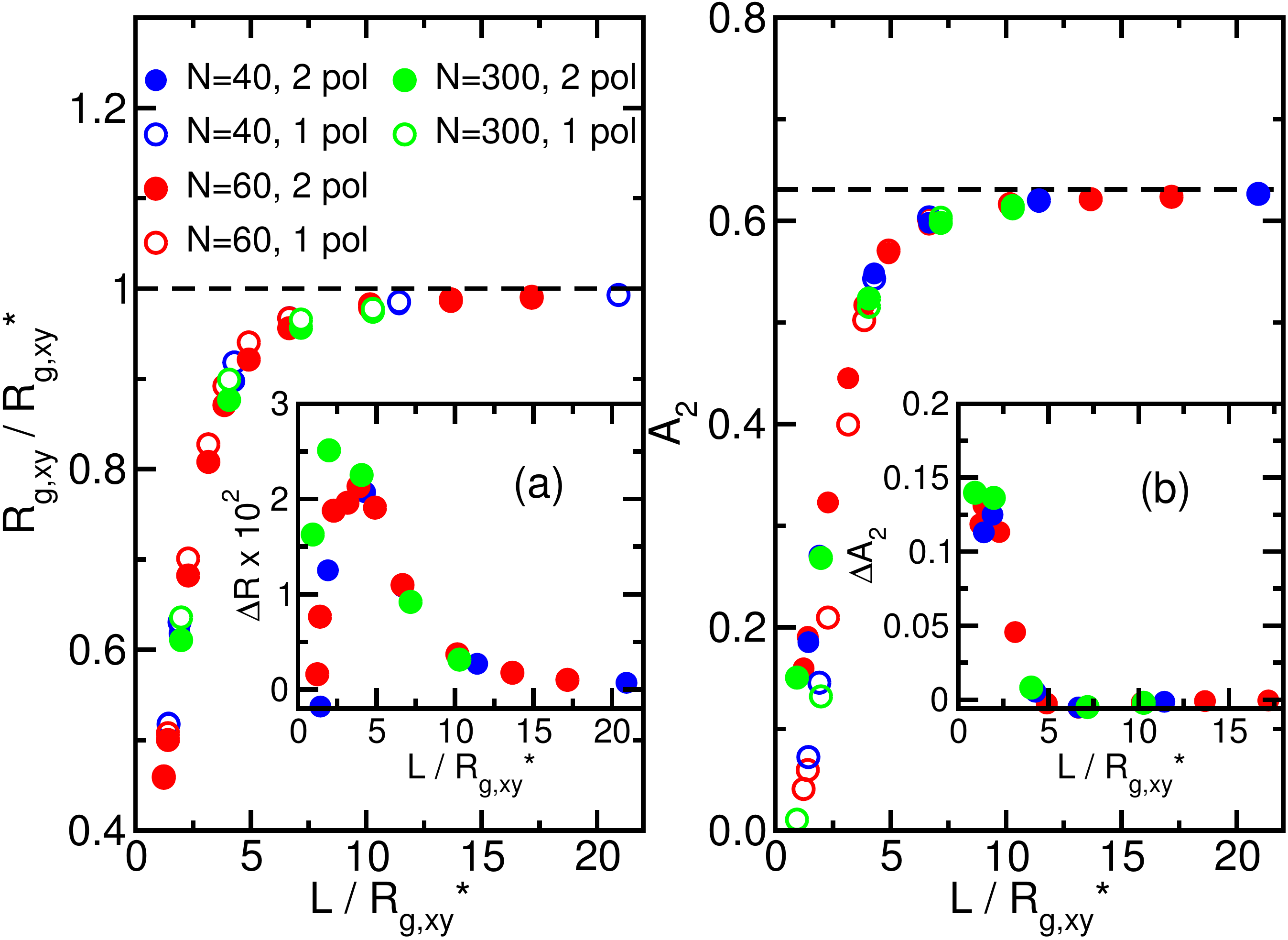}
\end{center}
\caption{(a) Scaled transverse radius of gyration, $R_{{\rm g},xy}/R_{{\rm g},xy}^*$,
vs. scaled box width, $L/R_{{\rm g},xy}^*$. Results are shown for box height $h$=4 and for both
1-polymer and 2-polymer systems for $N$=40, 60 and 300. The horizontal dashed line shows the value for 
$L$=$\infty$.  The inset shows the relative difference $\Delta R$ (defined in the text) between the 
data for the 1- and 2-polymer systems. (b) Asphericity, $A_2$, vs. scaled box width for the same systems 
as in panel (a).  The horizontal dashed line is the value measured for a slit for $N=60$. The inset 
shows the difference $\Delta A_2$ between the 1- and 2-polymer results.}
\label{fig:size_shape}
\end{figure}

Now let us examine conformational behaviour of each individual polymer.  Figure~\ref{fig:size_shape} 
illustrates the effect of lateral confinement on the mean size and shape of the polymer. 
Figure~\ref{fig:size_shape}(a) shows the variation of the transverse radius of gyration, 
$R_{{\rm g},xy}$, with respect to the box width. Results are shown for both 1- and 2-polymer 
systems for comparison. {For sufficiently large $L$, where polymer-wall and 
polymer-polymer interactions are infrequent} the transverse size of each polymer 
is close to the value for a slit For $L/R_{{\rm g},xy}^*\lesssim 5$, 
the size decreases rapidly with decreasing $L$. This decrease is comparable for both 1- and 2-polymer
systems, indicating that it is driven primarily by the interactions with the lateral confining walls
and less so by inter-polymer crowding. The relative difference between the two sets of results, 
$\Delta R \equiv (R_{{\rm g},xy}(1~{\rm pol})-R_{{\rm g},xy}(2~{\rm pol}))/R_{{\rm g},xy}^*$, is 
shown in the inset of Fig.~\ref{fig:size_shape}(a). The difference peaks at $L\approx 
4R_{{\rm g},xy}^*$, below which it decreases rapidly. Thus, at very small $L$, the 
inter-polymer crowding no longer drives compression of the polymer in the $x-y$ plane. This likely
arises from the increased tendency of the polymers to overlap with each other at the centre
of the box in this regime.

\begin{figure}[!ht]
\begin{center}
\vspace*{0.2in}
\includegraphics[width=0.4\textwidth]{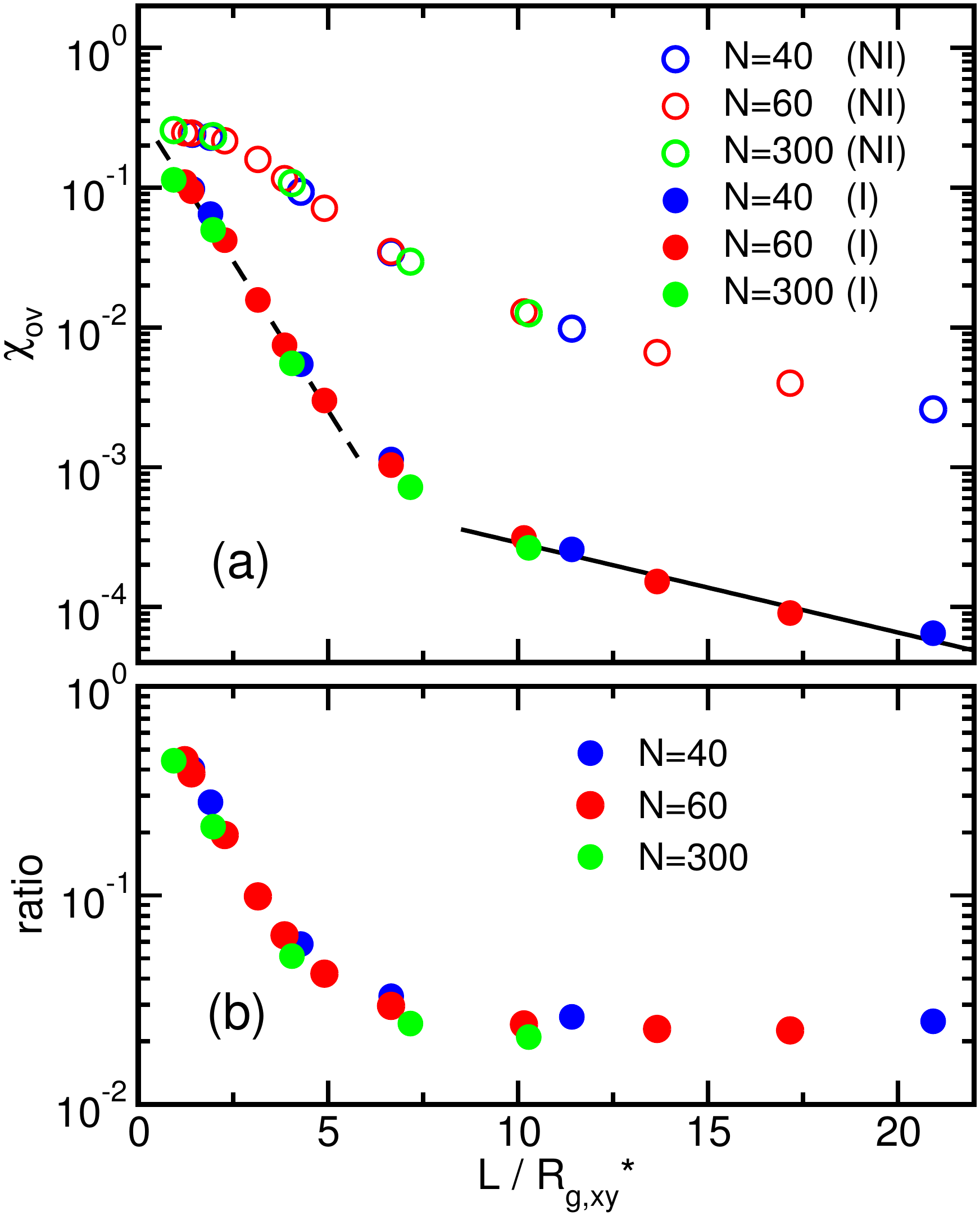}
\end{center}
\caption{(a) Variation of the overlap parameter, $\chi_{\rm ov}$, with scaled box width, 
$L/R_{{\rm g},xy}^*$. Results are shown for $N$=40, 60, and 300 for both interacting (I)
and noninteracting (NI) polymer systems. The definitions of these systems is described in
the text. The dashed and solid lines are fits to the interacting-system data in the
regimes $L/R_{{\rm g},xy}^*<5$ and $L/R_{{\rm g},xy}^*>10$, respectively. (b) Ratio of 
the overlap parameter for the {NI and I} cases as a function of box width.} 
\label{fig:overlap}
\end{figure}

Fig.~\ref{fig:size_shape}(b) shows the variation of the asphericity $A_2$, defined in 
Eq.~(\ref{eq:asph}) with respect to box width. As in Fig.~\ref{fig:size_shape}(a), the results
are comparable to those for a slit when $L$ is sufficiently large. As $L$ decreases, increased 
polymer-wall and polymer-polymer interactions result in a reduction in $A_2$. This implies that 
each polymer becomes less elongated and more disc-like as crowding increases.  The difference 
between the 1- and 2-polymer results is shown in the inset. While qualitatively similar to the data 
in the Fig.~\ref{fig:overlap}(a) inset for the polymer size, the difference for $\Delta A_2$ peaks 
at a much smaller box size of $L\approx 1.6R_{{\rm g},xy}^*$. The difference is appreciable.
For example, at $L$=7 we note $\Delta A_2=0.13$, which is 21\% of the asphericity for slit confinement.
In addition, maximum $\Delta A_2$ occurs at greatest confinement, precisely where the size difference 
is negligible and where Fig.~\ref{fig:prob2d_N60} indicates that the polymers in the 2-polymer
system have a tendency to overlap in the middle of opposite halves of the box. The concomitant 
inter-polymer repulsion evidently leads to a larger asphericity than for the case of a single 
polymer that interacts solely with the walls.

Given the effects of inter-polymer repulsion on polymer size and shape for small and medium
box widths, it is instructive to quantify the degree of polymer overlap. As described in 
Section~\ref{subsec:measurements}, {we define the overlap parameter}
$\chi_{\rm ov }\equiv N_{\rm ov}/N$, where $N_{\rm ov}$ is the mean number of monomers inside
overlapping equivalent ellipses for the polymers.
%
%
As a reference, we also show results for an {\it artificial} system of two 
{\it non}-interacting polymers, i.e., a model system in which overlap between pairs of monomers 
on different polymers are ignored (though non-bonded {\it intra}-polymer interactions are present).
We refer to this as a ``non-interacting'' (NI) system, and the real system as an ``interacting''
(I) system.

Figure~\ref{fig:overlap}(a) shows the variation of $\chi_{\rm ov }$ with box size for three 
different polymer lengths.  The degree of overlap is very small for large $L$ 
and increases significantly with increasing confinement. This is true for both interacting 
and non-interacting polymers. The values for the non-interacting system are larger than those 
for the interacting system. This arises simply from the entropic repulsion between polymers 
caused by the inter-polymer interactions. In the absence of such interactions the polymers 
have a greater tendency to overlap in the $x-y$ plane. For interacting polymers, the figure 
shows a transition in the rate of change of overlap with box width at $L/R_{{\rm g},xy}^*\approx 5$.
For $L/R_{{\rm g},xy}^*\lesssim 5$ the decrease in $\chi_{\rm ov}$ with box width is exponential
characterized by a decay constant of $\approx 1.0$, while for $L/R_{{\rm g},xy}^*\gtrsim 5$ 
the exponential decay constant is $\approx 6.8$. Thus, the degree of overlap increases rapidly
with increasing confinement in the regime for $L/R_{{\rm g},xy}^*\lesssim 5$. This inference 
is corroborated by the results in Fig.~\ref{fig:overlap}(b), which shows the ratio of the 1- 
and 2-polymer values for $\chi_{\rm ov }$. For large box widths of $L/R_{{\rm g},xy}^*\gtrsim 5$, 
each ratio is small and constant. For smaller box widths of $L/R_{{\rm g},xy}^*\lesssim 5$ the 
ratio rapidly increases for smaller box widths. Together, the results of Fig.~\ref{fig:overlap}(a) 
and (b) indicate that the entropic repulsion preventing overlap of interacting polymers is being 
overridden by the even stronger repulsion from the walls, contact with which becomes increasingly 
unavoidable at smaller $L$. This region of enhanced overlap coincides with that of the dramatic 
reduction in size and asphericity of the polymers. {Finally,} in Fig.~3 of the ESI, 
{results obtained using an alternative as a measure of overlap} show the same 
qualitative trend as in Fig.~\ref{fig:overlap}(b).

To summarize, the equilibrium statistics of the confined-polymer system exhibits behaviour
that depends strongly on the lateral width, $L$, of the confining box. At $L/R_{{\rm g},xy}^*\gtrsim 5$ 
the polymers rarely interact with each other or with the wall, and each chain behaves in most ways 
comparably to a single slit-confined polymer. For smaller boxes with 
$L/R_{{\rm g},xy}^*\lesssim 5$, the effects of confinement and inter-polymer crowding become 
appreciable. As $L$ decreases in this regime, the centres of mass of the two become increasingly 
localized at positions in opposite halves of the box, except for very small boxes where there
is a simultaneous tendency for them to overlap the the box centre. In addition, there is an
increase in the overlap in the lateral distributions of monomers, which is driven by the
increase in confinement and, counter-intuitively, enhanced by inter-polymer repulsion.
We also find that increasing lateral confinement decreases the average size
of the chains, an effect that is enhanced by inter-polymer crowding. The shape anisometry
(``asphericity'') also decreases with decreasing $L$ in this regime, though this effect
is slightly offset by that caused by inter-chain crowding.

\subsection{Polymer dynamics}
\label{subsec:dynamics}

We now examine the equilibrium dynamics of the confined-polymer system. As in Section~\ref{subsec:statics} 
we choose a single box height of $h$=4 and consider the effects of varying the box width. The trends 
in the dynamics can then be explained in the context of those for the equilibrium statistics described 
in the previous section.  Both the diffusion of the centre of mass and the internal dynamics 
are characterized.  As in Section~\ref{subsec:statics} it is useful to consider a 1-polymer system for 
comparison with the 2-polymer case in order to elucidate the effects of inter-polymer repulsion on 
the system behaviour.

\begin{figure}[!ht]
\begin{center}
\vspace*{0.2in}
\includegraphics[width=0.45\textwidth]{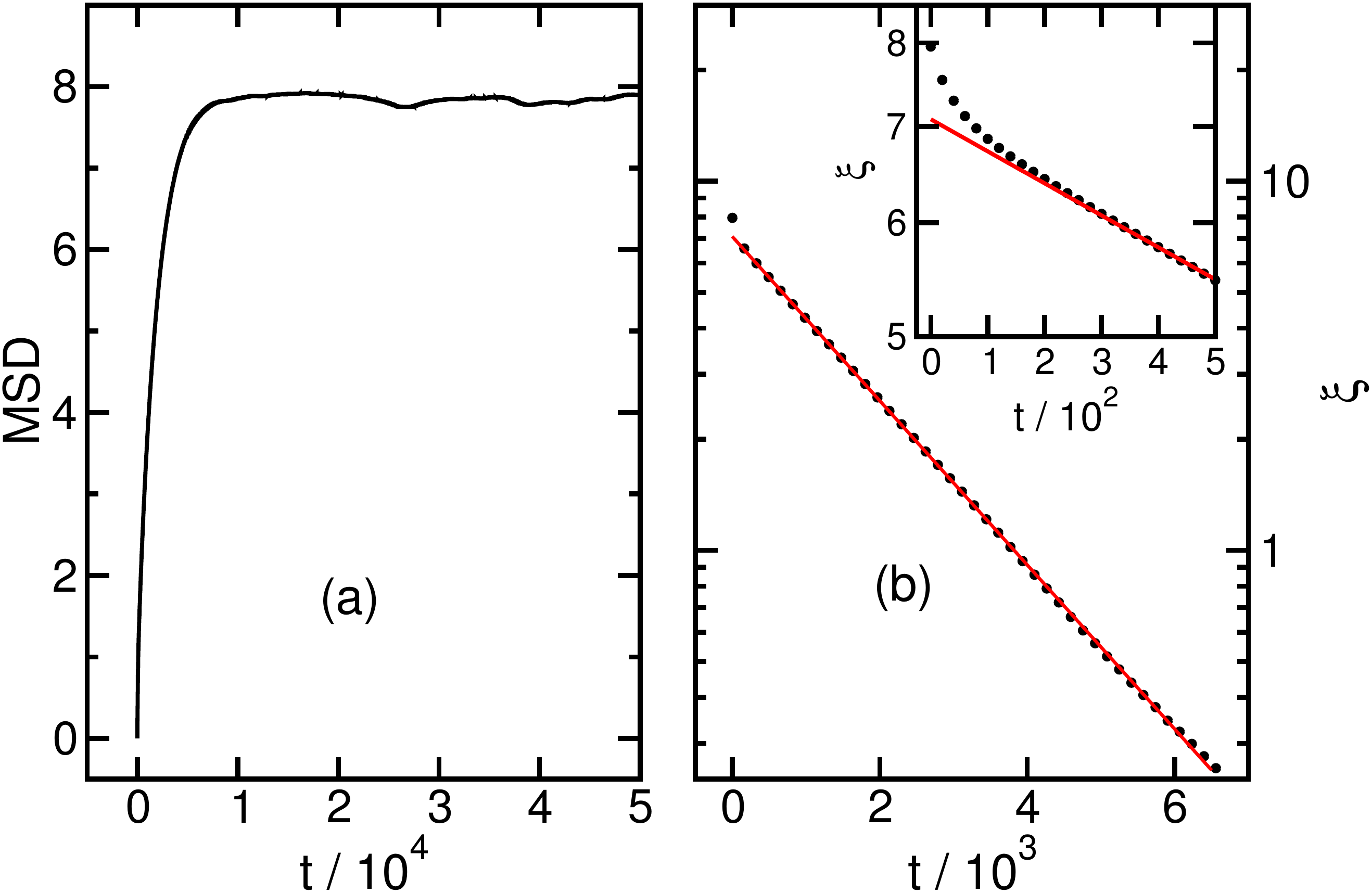}
\end{center}
\caption{(a) MSD versus time for a polymer of length $N$=60 in a box of width 
$L$=13 and height $h$=4. (b) $\xi$ versus time for the system in (a), where 
$\xi\equiv C_0 - {\rm MSD}(t)$, and where $C_0$ is the mean value of the MSD for $t>20000$.  
The red curve is a fit to an exponential using data in the range of $t=1000-6000$.
The inset is a closeup of the data at small $t$ illustrating the deviation in the
exponential fit in this regime.}
\label{fig:msd}
\end{figure}

We consider first the mean-square displacement (MSD) of the polymer centre of mass, defined
in Eq.~(\ref{eq:MSDdef}).
Figure~\ref{fig:msd}(a) shows the time dependence of the MSD for a 2-polymer system with $N$=60, 
$L$=13. Initially, the curve rises rapidly with time, after which it levels off to a 
constant value of approximately 7.8. The leveling off of the MSD is a straightforward consequence 
of the confinement of the polymer in the lateral directions. The basic features of the MSD shown 
in the figure are generic to the results for all $L$.  Generally, the limiting value of the MSD 
at long time and the characteristic time, $\tau$ (defined below), required to reach this plateau 
both increase with increasing box size.  Increasing the polymer length reduces the long-time 
value of the MSD and increases $\tau$.
 
A quantitative analysis of the MSD is aided by employing a theoretical model used in
Ref.~\onlinecite{capaldi2018probing} to analyze comparable experimental data. In this 
description, the centre-of-mass motion of a single DNA molecule confined to a box-like cavity is
modeled as free Brownian diffusion of a particle subject to an infinite square-well 
potential.\cite{kusumi1993confined} Interactions between polymer chains are effectively
ignored, implying that the model should only be quantitatively accurate for sufficiently wide boxes.
Since the cavity is symmetric in the $x$ and $y$ dimensions, the MSD is the same along 
these axes and can be averaged.  Solving the diffusion equation for a single particle 
in a square box of side length $L_{\rm e}$ yields\cite{kusumi1993confined}
\begin{eqnarray}
{\rm MSD}(t) = C_0 - C_1 \sum_{n=1,3,5,...}^\infty 
\frac{1}{n^4}\exp\left[-D\left(\frac{n\pi}{L_{\rm e}}\right)^2 t \right],~~
\label{eq:MSD}
\end{eqnarray}
where $D$ is the diffusion coefficient of the particle, and where $C_0\equiv L_{\rm e}^2/6$
and $C_1\equiv 16L_{\rm e}^2/\pi^4$. Defining the quantity $\xi(t)$ as
 \begin{eqnarray}
\xi(t) \equiv C_0 - {\rm MSD}(t),
\label{eq:xi}
\end{eqnarray}
and noting that all of the terms with $n\geq 3$ and are negligible in comparison to the $n=1$ 
term for sufficiently long $t$, it follows that
\begin{eqnarray}
\xi(t) \approx C_1 \exp\left(-t/\tau)\right)
\label{eq:xi2}
\end{eqnarray}
in this long-time limit, where $\tau\equiv DL_{\rm e}^2/\pi^2$.

To apply these results to the mean-square displacement of the centre of mass of a polymer 
diffusing in two dimensional square box, a reasonable choice of the effective box length
is $L_{\rm e}=L-2R_{{\rm g},xy}^*$, where $L$ is the true side length of the confining box
and $R_{{\rm g},xy}^*$ is the radius of gyration of the polymer measured in the $x-y$ plane
for a slit geometry. Thus, the polymer is modeled as a hard 2-D disk of radius $R_{{\rm g},xy}^*$, 
and shape deformations associated with pressing the polymer against a side wall are 
neglected.\cite{capaldi2018probing} In addition, noting that Rouse dynamics are obeyed for 
the simulation model, it follows that $D=k_{\rm B}T/\gamma$. Since the friction efficient 
satisfies $\gamma=N\gamma_0$, where $\gamma_0$ is the friction per monomer, and noting that 
$k_{\rm B}T/\gamma_0=1$, it follows that $D=1/N$. Consequently, the time constant in 
Eq.~(\ref{eq:xi2}) satisfies:
\begin{eqnarray}
\tau/N = L_{\rm e}^2/\pi^2.
\label{eq:tauN}
\end{eqnarray}

Figure~\ref{fig:msd}(b) shows the time dependence of the quantity $\xi$, calculated using the 
data in Fig.~\ref{fig:msd}(a).  {Consistent with} the theoretical model, we find 
that $\xi$ decreases exponentially with time. Only a small deviation from this behaviour is 
observed at low values of $t$, as illustrated in the inset of the figure. This arises from 
the non-negligible contribution to the MSD from the higher-$n$ terms in Eq.~(\ref{eq:MSD}).

\begin{figure}[!ht]
\begin{center}
\vspace*{0.2in}
\includegraphics[width=0.45\textwidth]{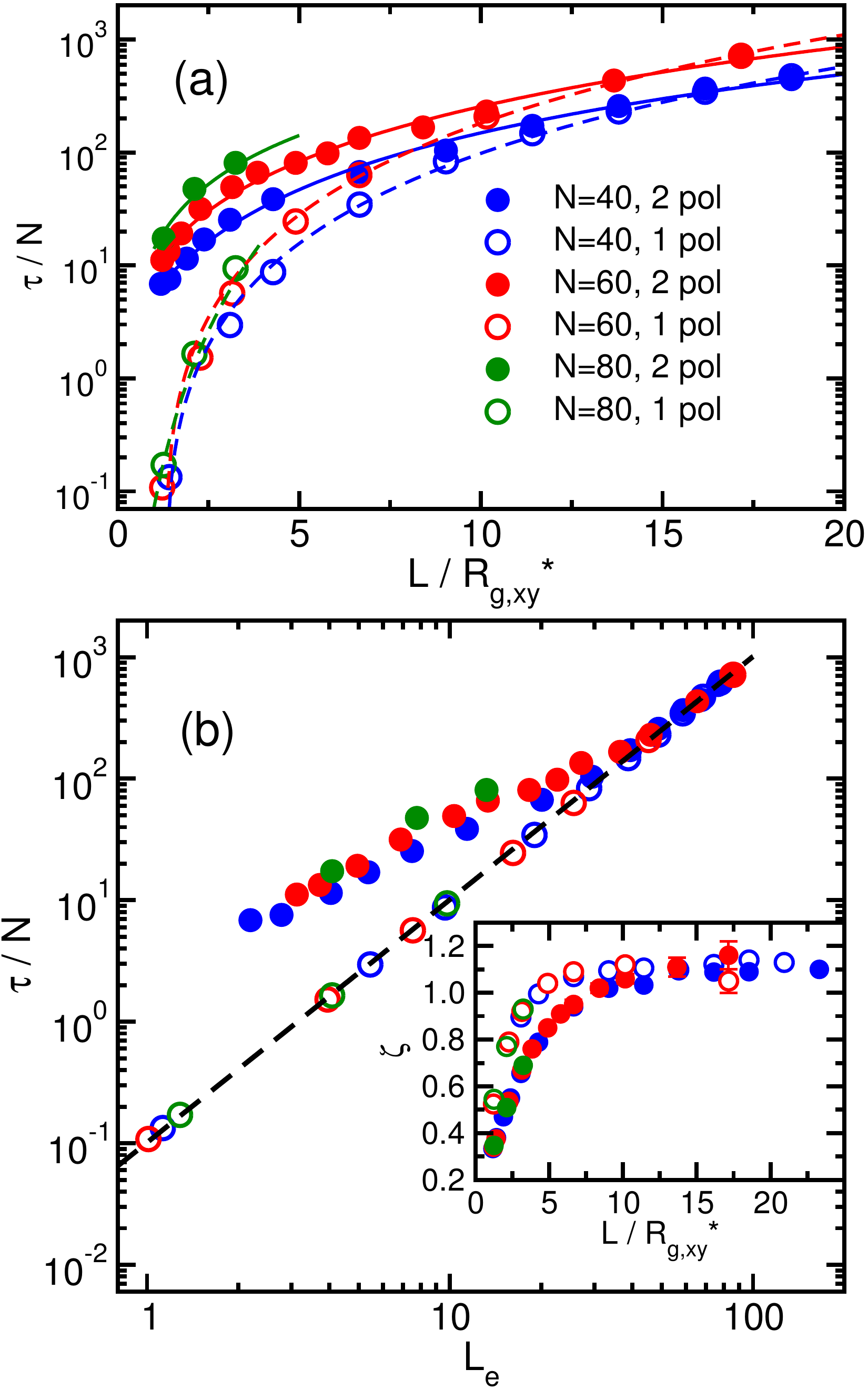}
\end{center}
\caption{(a) Variation of $\tau/N$ with respect to $L/R_{{\rm g},xy}^*$. The time constant $\tau$ 
is  extracted from fits to $\xi(t)$ at sufficiently long times, where $\xi(t)\equiv C_0-{\rm MSD}(t)$, 
and where $C_0 = \lim_{t\rightarrow\infty} {\rm MSD}(t)$. Results are shown for systems of a single
polymer (open symbols) and for two polymers (closed symbols) and for polymer lengths of
$N$=40, 60 and 80. The solid and dashed lines are guides for the eye for 2-polymer and 1-polymer
data sets, respectively.  (b) $\tau/N$ vs. $L_{\rm e}$, where the effective box width is defined
$L_{\rm e}=\sqrt{6C_0}$, as discussed in the text.  The dashed line shows the prediction of 
Eq.~(\ref{eq:tauN}), which is expected to be valid at $L_{\rm e}/R_{{\rm g},xy}^*\gg 1$.  
{The inset shows the variation of $\zeta\equiv (L-L_{\rm e})/2R_{{\rm g},xy}^*$ with 
$L/R_{{\rm g},xy}^*$.}  }
\label{fig:tau.N.Le}
\end{figure}

Figure~\ref{fig:tau.N.Le}(a) shows the variation of the scaled time constant, $\tau/N$, with
respect to the scaled box length, $L/R_{{\rm g},xy}^*$, while Fig.~\ref{fig:tau.N.Le}(b) shows
$\tau/N$ vs. the effective box length, $L_{\rm e}$. Here, $\tau$ is extracted from the fit of 
$\xi$ to Eq.~(\ref{eq:xi2}), and the effective box with is obtained from $L_{\rm e} = \sqrt{6C_0}$,
where the quantity $C_0$, defined in Eq.~(\ref{eq:MSD}), is estimated from the MSD in the 
plateau region.  Results are shown for both 1- and 2-polymer systems for polymer lengths of $N$=40, 
60 and 80. In Fig.~\ref{fig:tau.N.Le}(b) the theoretical prediction of Eq.~(\ref{eq:tauN}) 
is shown as a dashed curve. As expected, the data for both 1- and 2-polymer systems both 
converge to the theoretical curve in the limit of large box length. In this regime, interactions
between the two polymers are infrequent and therefore do not change the dynamical behaviour
of either polymer. As $L_{\rm e}$ decreases, such interactions become more frequent, leading
to a reduction in the rate of diffusion of the polymers. This is characterized by an increase
in the time constant relative to both the predictions of the theoretical model and the 1-polymer
time constant. Surprisingly, the prediction of Eq.~(\ref{eq:xi2}) remains very accurate
for the 1-polymer system even to very small box sizes where $L_{\rm eff}\approx R_{\rm g,xy}^*$.

As an additional test of the approximations employed in the theoretical model, we consider 
the quantity $\zeta \equiv (L-L_{\rm e})/2R_{{\rm g},xy}^*$.  As noted above, the effective 
box width is expected to be $L_{\rm e} \approx L-2R_{{\rm g},xy}^*$ if we neglect the deformations 
in the size and shape of the polymer. In this picture, $\zeta$ is a constant of 
order unity.  {The inset of Fig.~\ref{fig:tau.N.Le}(b) shows the measured variation of
$\zeta$ with $L/R_{{\rm g},xy}^*$ for both 1- and 2-polymer systems.} For the single-polymer
system, $\zeta$ is indeed constant and close to unity for $L/R_{{\rm g},xy}^* \gtrsim 5$. 
However, for $L/R_{{\rm g},xy}^* \lesssim 5$, $\zeta$ decreases slightly with decreasing box width.
In this regime, there is no extended area in the $x-y$ plane over which the polymer centre of mass 
can move without interacting with the walls. Thus, size and shape deformations are significant 
and neglecting them leads to the observed deviation from the prediction.

A similar trend is observed for the 2-polymer system. However, in this case the transition
occurs at a larger box width of $L/R_{{\rm g},xy}^*\approx 10$. The additional crowding
caused by the presence of the second polymer leads to increased interaction with the lateral
walls as well as shape deformations at box sizes where such effects are not as prominent
in the 1-polymer system.  Note that the 2-polymer $\zeta$ begins to decrease significantly 
with decreasing $L$ at $L/R_{{\rm g},xy}^*\approx 5$, which corresponds to $L\approx 29$ and
$L_{\rm e}\approx 17$. This is precisely where the measured $\tau/N$ for the 2-polymer system 
begins to deviate from the dilute-limit approximation. 

\begin{figure}[!ht]
\begin{center}
\vspace*{0.2in}
\includegraphics[width=0.4\textwidth]{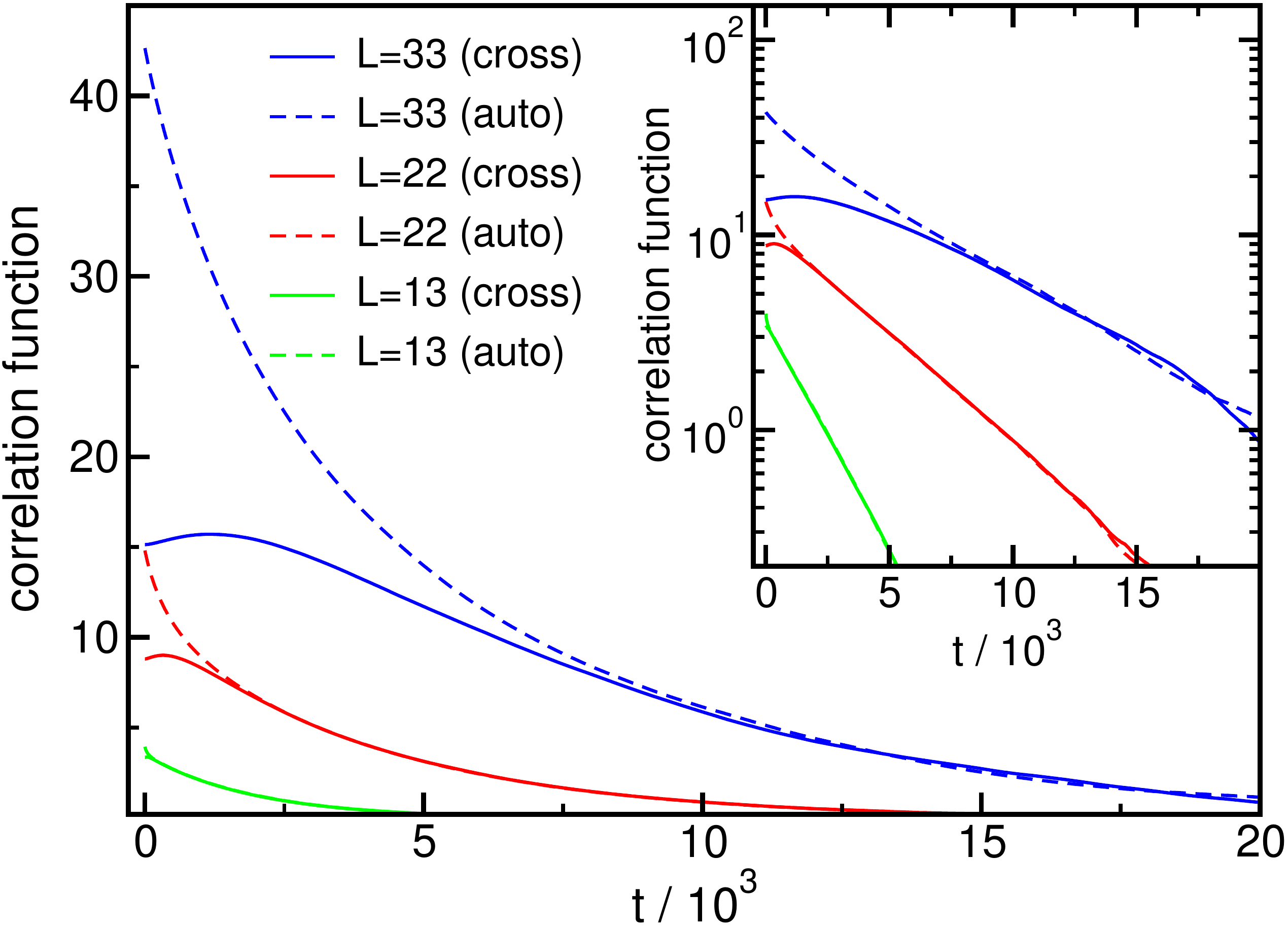}
\end{center}
\caption{Autocorrelation function, $C_{\rm auto}\equiv\left\langle x_1(t)x_1(0)\right\rangle$, and 
cross-correlation function, $C_{\rm cross}\equiv -\left\langle x_1(t)x_2(0)\right\rangle$, for the 
centre-of-mass positions for two $N$=60 polymers.  Results are shown for $L$=13, 22, and 33. The inset 
shows the same data plotted on a semi-log scale. }
\label{fig:corr}
\end{figure}

Next, we examine the time dependence of correlations in the centre-of-mass positions of
the polymers. We use the autocorrelation function, $C_{\rm auto}(t)$, and cross-correlation
function, $C_{\rm cross}(t)$, defined in Eqs.~(\ref{eq:Cauto}) and (\ref{eq:Ccross}), respectively. 
Figure~\ref{fig:corr} shows both functions for box widths of $L$=13, 22, and 33. Both correlation 
functions tend decay exponentially at large $t$. Note that in each case the decay to zero at long 
$t$ is a consequence of choosing $x=y=0$ at the box centre.  The individual functions diverge with 
decreasing $t$. This divergence is greater for larger box sizes.

The system behaviour at $t=0$ has a simple explanation. Note that $C_{\rm auto}(0)=\langle x_1^2 \rangle$
and $C_{\rm cross}(0)=-\langle x_1 x_2\rangle$. As previously noted for the results in Fig.~\ref{fig:corr_LJ},
$\langle x_1^2 \rangle$ approaches $-\langle x_1x_2 \rangle$ for small box sizes since the polymer positions 
in the $x-y$ plane are highly anti-correlated in this regime (i.e. if one polymer lies at $(x,y)$ the other 
is likely to be near $(-x,-y)$).  For larger box sizes the degree of anticorrelation is smaller, leading 
the observed wider divergence between the correlation functions for larger $L$. 

To better understand the origins of the observed time dependence of the functions, we employ a 
simple model to describe the box-size regime where the polymers are pushed out from the centre of 
the box (see, e.g., the results for $L$=13 and $L$=22 in Fig.~\ref{fig:prob2d_N60}). Here, the 
polymers are pictured as two point particles that each occupy one of four discrete sites on the 
corners of a square and whose interactions mimic the effects of entropic repulsion. The model can 
be used in MC dynamics simulations to measure the two position correlation functions. The model 
is fully described in the appendix, and the calculated correlation functions are shown 
in Fig.~\ref{fig:corr_discrete_model}. The convergence to exponential decay at large $t$ and the 
divergence at low $t$ are both present. Thus, a simple model incorporating the basic features of 
the probability distributions of Fig.~\ref{fig:prob2d_N60} and a simple description of entropic 
repulsion between the polymers is capable of accounting for the general behaviour of the correlation 
functions. 

Let us now examine the effects of confinement on the internal motion of the polymers.
A useful means to characterize such motion is the autocorrelation function of the
Rouse coordinates of the polymer. As noted in Section~\ref{sec:methods}, these functions
tend to be exponential over at least one decade of decay (except at very short times)
for both 1- and 2-polymer systems.  

\begin{figure}[!ht]
\begin{center}
\vspace*{0.2in}
\includegraphics[width=0.45\textwidth]{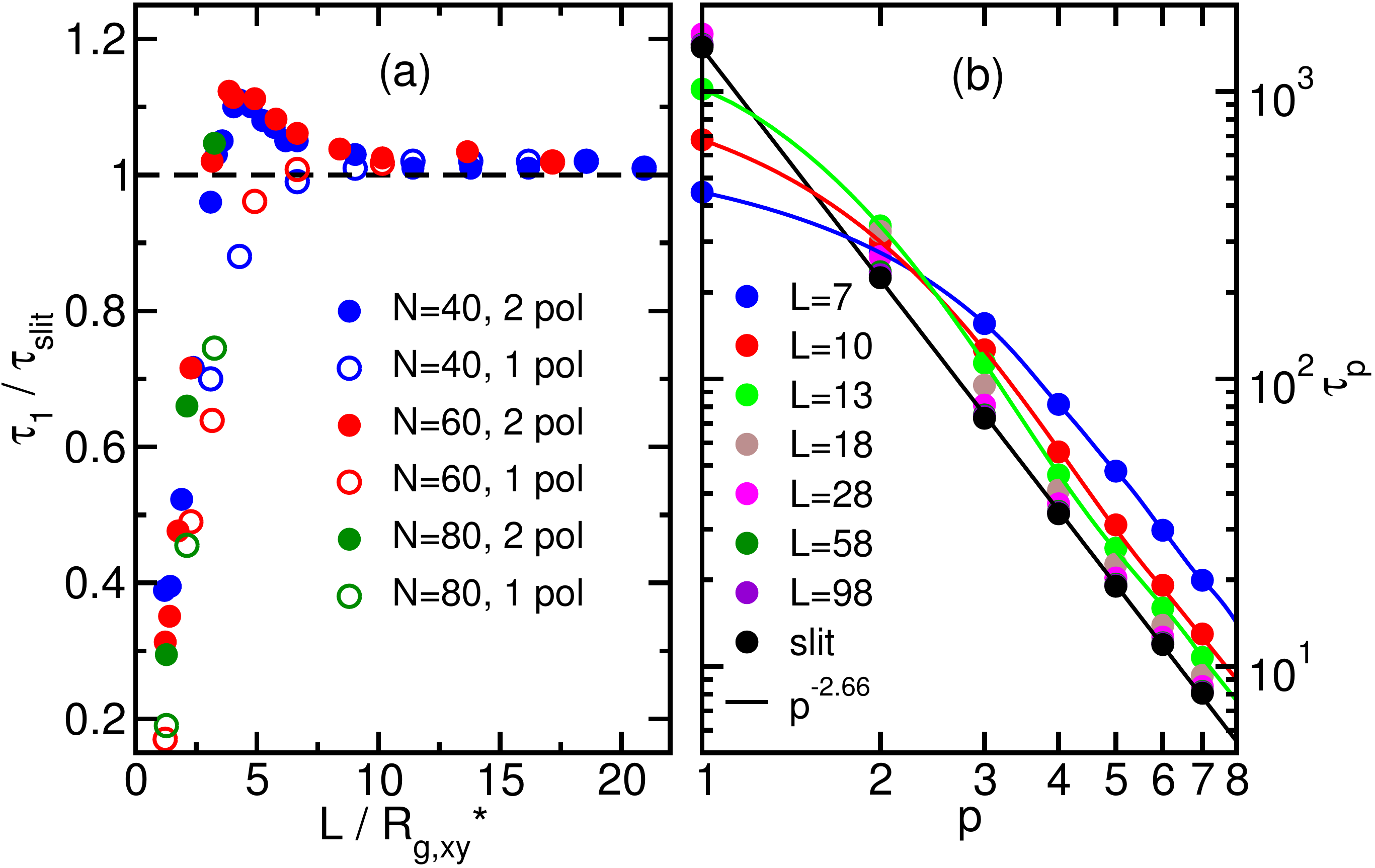}
\end{center}
\caption{(a) Variation of scaled Rouse mode correlation time $\tau_1/\tau_{\rm slit}$ with box 
width, $L$, where $\tau_{\rm slit}$ is the value of $\tau_1$ for slit confinement. Results are shown
for $N$=60 for 1- and 2-polymer systems. The dashed line corresponds to $\tau_1=\tau_{\rm slit}$.  
(b) Variation of $\tau_p$ with mode index $p$ for a 2-polymer system with $N$=60. Results are shown 
for various box widths. The black curve shows a fit to the results for slit-confinement ($L=\infty$) 
to a power law.  The other curves are guides for the eye for $L$=6, 8, and 13. }
\label{fig:rouse}
\end{figure}

Figure~\ref{fig:rouse}(a) shows the variation of the Rouse mode correlation times for the 
$p$=1 mode with respect to box width. Note that $\tau_1$ describes rates of conformational 
motion on large length scales.  For sufficiently wide boxes $\tau_1$ approaches 
the value for slit confinement for both 1- and 2-polymer systems. Thus, the rate
of conformational change is independent of box width when the polymers rarely encounter
the confining walls.  However, as $L$ decreases, the behaviours of the two systems diverge.
For the 2-polymer system the $\tau_1$ exhibits a peak at at $L/R_{{\rm g},xy}^*\approx 4$, 
followed by a sharp decrease at lower $L$.  By contrast, the 1-polymer system has no maximum 
in this region, and instead $\tau_1$ exhibits solely a sharp decrease at 
$L/R_{{\rm g},xy}^*\lesssim 5$.

The effect of confinement on Rouse mode relaxation has been previously examined 
analytically using a Gaussian chain subject to a harmonic confining potential.\cite{denissov2002segment}
In each dimension subject to this confinement, it was shown that $C_p(t)$ decays exponentially
with a relaxation time
\begin{eqnarray}
1/\tau_p = 1/\tau_p^{(0)} + 1/\tau_{\rm e},
\label{eq:taupinv}
\end{eqnarray}
where $\tau_p^{(0)}\propto (N/p)^2$ is the relaxation time for an unconfined chain,
and $\tau_{\rm e}$ is a constant proportional to $\tilde{d}^2$, where $\tilde{d}$ is the
effective confinement width. Thus, in this model $\tau_p$ decreases as the confinement
increases, qualitatively consistent with the present results for $\tau_1$ for a single 
self-avoiding chain laterally confined between hard walls. The presence of a second 
polymer chain effects an increase in $\tau_1$ over the range of $L$ where the two polymers
are forced to be in contact with each other, including lower values of $L$ where 
$\tau_1<\tau_{\rm slit}$. This is likely an effect of polymer crowding impeding conformational 
motion on large length scales. The competition between the effects of confinement, which
tends to decrease $\tau_1$, and crowding, which tends to increase $\tau_1$, leads to
the local maximum. 

Figure~\ref{fig:rouse}(b) shows the variation of the Rouse mode correlation times, $\tau_p$,
as a function of mode index $p$ for the 2-polymer system with $N$=60. Results are shown for various
box widths for mode indices in the range $p=1-7$. In all cases $\tau_p$ decreases monotonically
with $p$.  For confinement to a slit, i.e. $L=\infty$, the results exhibit power-law behaviour
with $\tau_p \propto p^{-2.66}$. This is close to the expected Rouse scaling for a 2-D polymer
in good solvent conditions: $\tau_p \propto p^{-(1+2\nu)} = p^{-2.5}$, where $\nu=0.75$ is the Flory
scaling exponent for two dimensions.\cite{deGennes_book} The discrepancy is likely a result
of finite-size effects. The relaxation times $\tau_p$ are independent of $L$ for $L \gtrsim 20$.
For $L\lesssim 20$ ($L/R_{{\rm g},xy}^*\lesssim 3.5$), $\tau_p$ increases with decreasing box width 
for all $p$ except for the case of $p=1$ at very small $L$, as noted above. Thus, in the confinement 
regime where 2-polymer crowding effects become noticeable, the conformational dynamics appear to 
be slowed over all length scales. Finally, we speculate that $\tau_p<\tau_p^{\rm (slit)}$ for 
$p\geq 2$ for smaller box widths than those examined in our simulations. This follows from the 
fact that Eq.~(\ref{eq:taupinv}) suggests that as $p$ increases the second term in 
Eq.~(\ref{eq:taupinv}) can remain appreciable relative to the first term if the confinement 
dimension $\tilde{d}$ (analogous to $L$) is reduced. 

To summarize, we find that the dynamics of the confined-polymer system are significantly affected 
by size of the confining cavity for sufficiently small $L$.  The centre-of-mass motion and the 
conformational dynamics are both influenced by a combination of confinement effects that are 
also present for a single polymer, as well as crowding effects arising from interactions between
the two polymers. These effects become pronounced precisely in the regime where the centre-of-mass 
probability distributions, the average size and shape of individual polymers, and the degree of 
polymer overlap are also strongly altered.

\section{Relevance to experiment}
\label{sec:experiment}

As noted in the introduction, this study is principally motivated by the recent work of Capaldi 
{\it et al.},\cite{capaldi2018probing} who used optical imaging methods to probe the organization 
and dynamics of DNA molecules trapped in nanofluidic cavities. Consequently, it is of value to 
examine the relevance of the results of the present study to those experiments. 

It is important to first note how the choice of model limits direct quantitative comparison. 
In their experiments using fluorescently stained $\lambda$ DNA, Capaldi {\it et al.} used a solution 
containing 10 mM tris at pH 8 with 2\% BME, corresponding to an ionic strength of about $I=$12~mM. 
Using the empirical relation between the Kuhn length, $l_{\rm K}$, and ionic strength from 
Dobrynin\cite{dobrynin2006effect} we estimate $l_{\rm K}\approx 100$~nm. Likewise,
using the theory of Stigter\cite{stigter1977interactions} for the relation between the $I$ and
the effective chain width, $w$, we estimate $w\approx 10$~nm, and thus the monomer anisotropy ratio
is $l_{\rm K}/w\approx 10$. For the staining ratio of 10:1 (bp:fluorophore) employed, the Kuhn 
length is not expected to change.\cite{kundukad2014effect} On the other hand, the contour length 
is expected to increase due to unwinding of the double helix. From Fig.~7 of 
Ref.~\onlinecite{kundukad2014effect} we estimate that a 10:1 YOYO-1 staining ratio leads to
a contour length increase of about 15\%. As unstained $\lambda$ DNA has a contour length
of $L_{\rm c}=16.5$~$\mu$m, the value for stained DNA is expected to be $L_{\rm c}\approx 19$~$\mu$m.
Thus, we estimate a ratio of $L_{\rm c}/l_{\rm K}\approx 190$.  The simulation model employed a 
flexible chain of spherical beads. If the bead width (which is approximately the mean bond 
length) represents one Kuhn length, then the model polymers should have a length of $N$=190 to 
achieve the correct $L_{\rm c}/l_{\rm K}$ ratio. This is larger by a factor of 2.4--5 than the 
polymer lengths ($N$=40--80) used in most simulations, with the exception of some calculations 
for $N$=300 carried out in MC simulations. More problematic is the ratio of $l_{\rm K}/w=1$ 
in the simulation model.  Finally, the ratio of the bulk radius gyration ($\approx 
700$~nm\cite{lin2011partial}) to the nanofluidic cavity depth (200~nm) is $R_{\rm g}/h\approx 
3.5$. By contrast, the ratio for the simulation model using $N$=60 and $h$=4 is $R_{\rm g}/h\approx 
1.35$. Thus, confinement in the narrow dimension of the cavity deforms $\lambda$-DNA significantly 
more than is the case in the model system. Increasing the ratio by decreasing $h$ in the simulation 
leads to small ratios of $h/w$ that result in unacceptable artifacts.  Choosing model parameter 
values to better match the other length scale ratios to those in the experiments
leads to simulations that require infeasibly long simulation times, especially in the case of 
Brownian dynamics simulations. Consequently, we must use a model system for which we can
expect only qualitative or, at best, semi-quantitative agreement between experiment and simulation.

The simulations examined cavities with a wide range of width values. By contrast the nanocavities
employed in the experiments had a single fixed width of 2~$\mu$m. The ratio of the in-plane 
radius of gyration (0.91~$\mu$m) for a slit confined $\lambda$-DNA molecule and box length 
was $L/R_{{\rm g},xy}^*\approx 2.2$. Imposing this ratio on the simulation model with $N$=60 
(for which $R_{{\rm g},xy}^*\approx 5.7$) implies a box width of $L\approx 13$.  
Figure~\ref{fig:prob2d_N60} shows that a single molecule confined to a box of this 
width is expected to have its centre of mass be strongly localized to the centre of the box. 
This is qualitatively consistent with the experimental results presented in Fig.~4(a) and 4(c) 
of Ref.~\onlinecite{capaldi2018probing}. The wider, more square-like distribution for 
the larger box sizes in Fig.~\ref{fig:prob2d_N60}(a) better resemble the
measured position distribution for confinement of a single plasmid shown in Fig.~5(a)
of Ref.~\onlinecite{capaldi2018probing}. Since the plasmid contour length was considerably 
shorter than that of the $\lambda$ DNA chain and was of circular topology, its average 
size was also much smaller. This naturally results in a distribution better approximated 
by a model system using a larger $L/R_{{\rm g},xy}^*$ ratio.

In the case of two confined polymers, the results for 
$L$=13 in Fig.~\ref{fig:prob2d_N60} suggest that crowding effects cause the polymer centres 
to be pushed out from the centre to opposite halves of the box at four quasi-discrete 
locations and leaving a probability ``hole'' at the box centre. The corresponding experimental
results of Fig.~4(d) of Ref.~\onlinecite{capaldi2018probing} do indicate a crowding-induced
displacement of the $\lambda$-DNA molecules from the box centre. However, in that case
an asymmetry was noted, likely a result of an underlying difference in the DNA contour and 
persistence lengths caused by using different stains, as required for separate observation of 
each molecule. Specifically, the YOYO-3 labeled DNA chain was slightly more concentrated in 
the cavity centre than the YOYO-1 labeled chain. In addition, the distribution of sampled
centre-of-mass positions shown in Fig.~4(b) of Ref.~\onlinecite{capaldi2018probing}
do not show any evidence of the quasi-discrete states, {which likely
arise from enhanced packing effects due to the small value of $l_{\rm K}/w$ ratio in
the model.} Instead, those results are more
qualitatively consistent with our simulation results for a slightly larger box width ($L$=22),
where ${\cal P}_1$ tends to be dependent on radial distance from the box centre alone,
independent of the polar angle, and where the probability hole at the centre is less pronounced. 
Such a distribution is also consistent with a collective Brownian rotation of the two molecules 
around the centre of the box, a behaviour noted in Ref.~\onlinecite{capaldi2018probing}.

In addition to measurement of the time-dependence of the MSD, Capaldi {\it et al.} also measured 
the position autocorrelation function, $C_{\rm auto}(t)$, and observed exponential decay for both
1- and 2-polymer systems. They found a time constant of $\tau=0.25\pm 0.01$~s for 1-chain confinement
and $\tau=2.0\pm 0.1$~s for 2-chain confinement. Thus, the crowding effect caused by the presence
of the second chain increased $\tau$ by a factor of 8.  It is easy to show that an exponential 
decay of the correlation function implies an exponential decay of the MSD with exactly the same 
time constant. Figure~\ref{fig:tau.N.Le} shows that the values of the time constant extracted 
from fits to the MSD diverge for the two different systems as $L_{\rm e}$ decreases. This
divergence begins at an effective box length of $L_{\rm e}\approx 20$.
As noted above, the cavity width in the experiment satisfies $L/R_{{\rm g},xy}^*\approx 2.2$.
For the model system, Fig.~\ref{fig:tau.N.Le}(a) shows that this ratio corresponds to the regime
where the presence of a second polymer increases the time constant relative to the
case for single-polymer confinement. Using the data in this figure we estimate an increase
in the time constant by a factor of 21 for $N$=60. {While this is larger than 
the experimental value, the model does correctly predict an increase in $\tau$ by about an
order of magnitude for the two-polymer system. Given the simplicity of the molecular model,
this is a reasonably good agreement.}

A final point of comparison with the experiments is the relationship between the center-of-mass
auto- and cross-correlation functions. Capaldi {\it et al.} reported exponential decay of
both $C_{\rm auto}(t)$ and $C_{\rm cross}(t)$ with time constants of $2.0\pm 0.1$~s
and $2.8\pm 0.3$~s, respectively. This stands in contrast to the simulation results
in which exponential decay only occurs at longer time, where the two functions converge
and thus are characterized by the same time constant. We suspect that this discrepancy
arises from the way in which the fit to the data was carried out in Ref.~\onlinecite{capaldi2018probing}.
The cross-correlation function in Fig.~7(c) of Ref.~\onlinecite{capaldi2018probing} 
appears to show a flattening of the curve at low $t$.  This trend is consistent with our 
simulation results, suggesting that it has a physical origin and is not merely a statistical
anomaly. Discarding the low-$t$ experimental data will likely increase the measured rate of decay 
and thus decrease the time constant. We speculate that such a modified fit could lead to a time
constant that better matches that for the autocorrelation function.

\section{Conclusions}
\label{sec:conclusions}

In this study we have used MC and Brownian dynamics simulations to probe the organization,
conformational behaviour, and equilibrium dynamics of two polymers under confinement in
a box-like cavity with very strong confinement in one of the dimensions.  {We find} 
the behaviour {is} highly dependent on the degree of lateral confinement. For large
box width, $L$, where the polymers rarely interact with each other or the lateral walls,
the polymer conformational statistics and dynamics are comparable to those for 
a single slit-confined polymer. The combined effects of confinement and inter-polymer
crowding are noticeable when $L/R_{{\rm g},xy}^*\lesssim 5$, where $R_{{\rm g},xy}^*$
is the transverse radius of gyration for the case $L\rightarrow\infty$ (i.e., confinement
to a slit with a spacing equal to the height of the confining cavity).
In this box size regime, there is a probability hole at the box centre, and the polymer 
centre-of-mass positions tend to be inversely correlated with respect to the box centre 
(i.e. when one polymer is at position $(x,y)$ the other tends to be at $(-x,-y)$, where
the box centre is at $(0,0)$). 
At lower $L$, the polymers tend to occupy four quasi-discrete states in opposite
sides of the box, and at very small $L$ there is an increasing tendency for polymer overlap
at the box centre. The polymer size decreases with $L$ in this regime, principally as a
consequence of confinement rather than interpolymer crowding. Increasing confinement tends
to reduce the 2D asphericity, though the interpolymer crowding tends to offset this effect.
Both the rate of diffusion and the internal dynamics of each polymer is significantly
impacted by the presence of the other polymer. {Note that the transition
location of $L/R_{{\rm g},xy}^*\approx 5$ is likely specific to the model employed in
this study. For other models (e.g. one using semiflexible chains) the transition
will likely occur at a somewhat different location, though still with $L/R_{{\rm g},xy}^*$
somewhat greater than unity.}

The simple molecular model employed in this study facilitates the study of a number of 
generic effects of confinement on the organization and dynamics of two cavity-confined polymers.
Generally, the observed behaviour is qualitatively consistent with the results of the recent 
experimental study by Capaldi {\it et al.}, which examined two $\lambda$-DNA chains confined 
to a nanofluidic cavity.\cite{capaldi2018probing} We view the present study as a first step 
toward a more realistic modeling of such experimental systems. In principle, scaling up the 
polymer length and incorporating bending rigidity into the model can be used to obtain correct 
lengthscale ratios for the contour length, persistence length, effective width of $\lambda$ DNA,
and the cavity dimensions. In practice, however, the required simulation times for such a model 
are infeasible at present, at least for the dynamics. A promising alternative approach could be 
to model each polymer as a chain of blobs with diameter equal to the box height, each interacting 
with other blobs via a soft repulsive potential arising from entropic repulsion. This potential 
could be determined using a technique to measure free energy functions employed in other recent
studies.\cite{polson2014polymer,polson2018segregation,polson2021free}
Such effective potentials were recently employed to model the entropic repulsion of side-loops
in a model for a bacterial chromosome.\cite{wu2019cell,swain2019confinement}
We anticipate that such future studies in addition to the present one will be helpful 
in elucidating recent experimental results as well motivating new nanofluidics experiments for 
cavity-confined DNA systems.

\appendix
\section{Position correlation functions for a discrete-site model }
\label{app:a}

To better understand the behaviour of the position correlation functions shown
in Fig.~\ref{fig:corr}, we carry out MC dynamics simulations using a discrete-site toy 
model that incorporates only the most basic features of the polymer system. 
The two polymers are modeled as interacting particles that can each occupy one of 
four discrete sites on the corners of a square in the $x-y$ plane. We are most interested
in the behaviour for intermediate box widths that correspond roughly to the experimental
regime. Here, the polymer centres are pushed out from the box centre in opposite directions
by roughly the same distance (e.g. see probability distributions for $L$=13 and $L$=22 in 
Fig.~\ref{fig:prob2d_N60}(b)). As a very simple approximation to that picture, we choose the
four site positions to be $(1,0)$, $(0,1)$, $(-1,0)$, and $(0,-1)$. The particles interact with
an energy $E\geq 0$ that is a measure of the entropic repulsion of the polymers. 
The interaction energy is $E=0$ when the particles are at opposite corners of the square, 
$E=E_1$ ($>0$) when the particles are on neighbouring corners, and $E=E_2$ ($>E_1$) when 
the particles occupy the same site. Thus, the energy decreases in strength with increasing
distance, in accord with the expected behaviour for entropic repulsion between polymers.

We perform a MC dynamics simulation on this system as follows. Each trial move consists 
of displacing a randomly chosen particle to either one of the two neighbouring sites. 
Each of these two neighbouring sites has equal probability of being selected. The trial
move is accepted or rejected according to the Metropolis criterion. The simulation is
carried out for $10^6$ MC cycles, where each MC cycle consists of two trial moves.
The position auto- and cross-correlation functions defined in Section~\ref{subsec:dynamics}
are calculated using the particle coordinates, which are sampled each MC cycle.
Time is measured in MC cycles.

\begin{figure}[!ht]
\begin{center}
\vspace*{0.2in}
\includegraphics[width=0.4\textwidth]{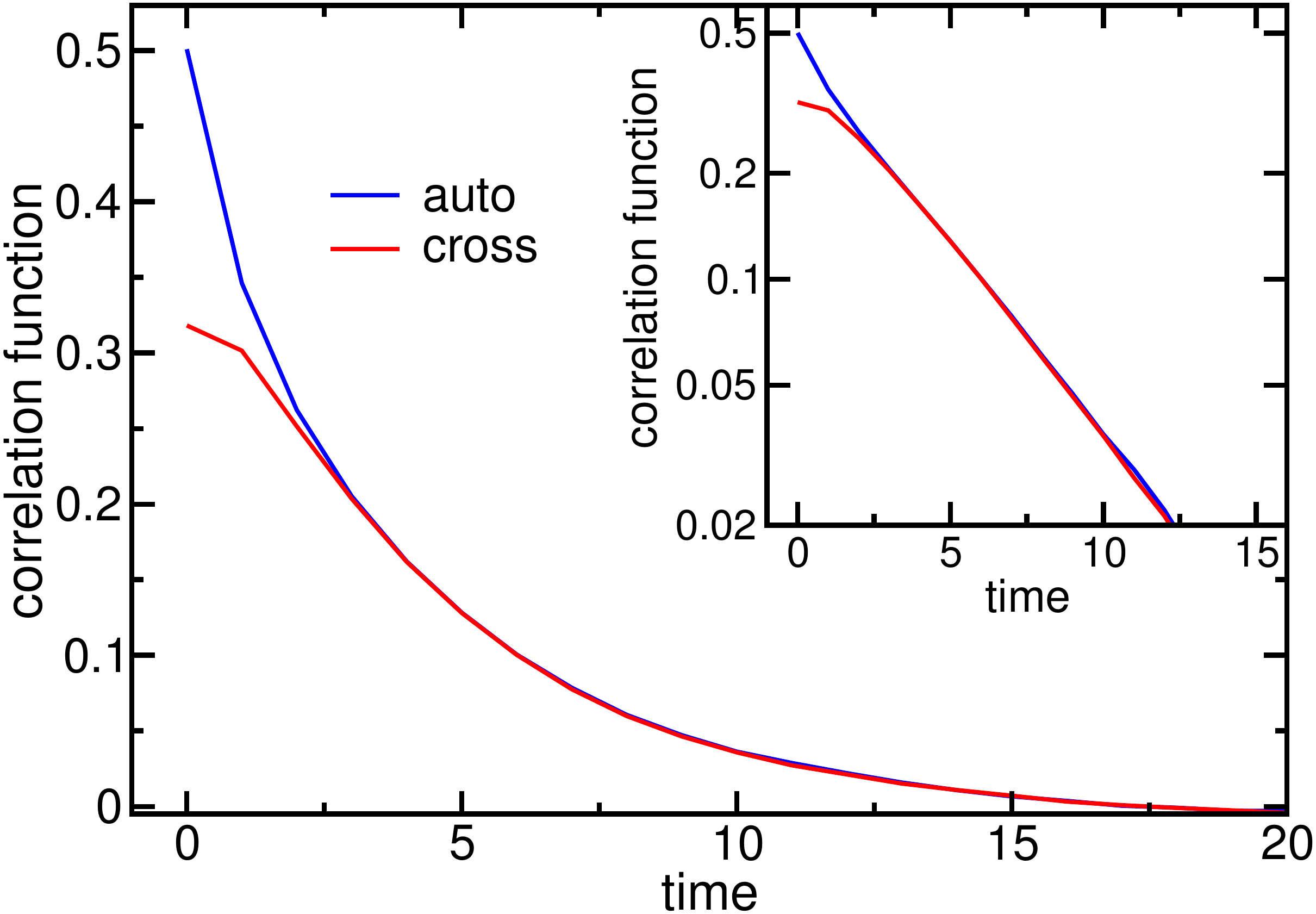}
\end{center}
\caption{Position auto-correlation and cross-correlation functions for the discrete-site
model described in the text of the appendix. The inset shows the same data plotted on a
semi-log scale.} 
\label{fig:corr_discrete_model}
\end{figure}

Figure~\ref{fig:corr_discrete_model} shows the position correlation functions 
using interaction energies of $E_1/k_{\rm B}T = 1$ and $E_2/k_{\rm B}T = 2$. As was
the case for the correlation functions for a system of two confined polymers shown in
Fig.~\ref{fig:corr}, $C_{\rm auto}(t)$ and $C_{\rm cross}(t)$ converge and decay exponentially
to zero at long times. Also consistent is the behaviour at short times, where the functions
diverge with $C_{\rm cross}(t)<C_{\rm auto}(t)$.

Qualitatively similar behaviour is observed for other choices of $E_1$ and $E_2$ ($>E_1$).
Decreasing the interactions energies corresponds to decreasing the entropic repulsion 
between polymers as a result of increasing the box size, $L$. Generally, we find that 
the divergence between the $C_{\rm auto}(t)$ and $C_{\rm cross}(t)$ at low $t$ is larger
when the repulsion is weaker, consistent with the trend for increasing $L$ in Fig.~\ref{fig:corr}.
The results presented in this appendix demonstrate that the general behaviour of the 
correlation functions can be accounted for using a model incorporating the qualitative features
of the probability distributions of Fig.~\ref{fig:prob2d_N60} as well as a simple description of
entropic repulsion between the polymers.

\begin{acknowledgments}
This work was supported by the Natural Sciences and Engineering Research Council of Canada (NSERC).
We are grateful to Compute Canada for use of their computational resources. DR would like to thank
Zhezhou Liu for helpful discussions. We would also like to thank Alex Klotz for helpful discussions
and for a critical reading of the manuscript.
\end{acknowledgments}


%

\end{document}